# Interatomic potential for the compound-forming Li-Pb liquid alloy

Alberto Fraile[a*], Santiago Cuesta-López[b,], Alfredo Caro[c], Daniel Schwen[c]
and J. Manuel Perlado[a]

[a]*Instituto de Fision Nuclear, ETSI Industriales, Universidad Politecnica de Madrid. Jose Gutierrez Abascal, 2 28006, Madrid, Spain.*
[b]*Universidad de Burgos. Parque Científico I+D+I. Plaza Misael Bañuelos s/n, 09002, Burgos, Spain*
[c]*Los Alamos National Laboratory, Los Alamos NM 87545 USA*

**Abstract**

Atomistic simulations of liquid alloys face the challenge of correctly modeling basic thermodynamic properties. In this work we present an interatomic potential for the Li-Pb system, as well as a study of physical properties of Li-Pb alloys. Despite the complexity due to Li-Pb being a compound forming system where charge transfer is expected, we show here how the empirical EAM formalism is able to satisfactorily describe several physical properties in a wide range of Li concentration. Application of our potential to Li-Pb eutectic allows us to correctly predict many physical properties observed experimentally and calculated with *ab initio* techniques, providing in this way a potential suitable for future studies in the context of tritium breeder blanket designs in Fusion technology.

**1. Introduction**

One of the main issues in current Nuclear Fusion programs is the design of liquid metals breeder blankets. In different blanket concepts, a eutectic of lead and lithium (Li17Pb) is foreseen as tritium breeder [1, 2]. Nowadays, understanding the structure and dynamic behaviour of liquid metal alloys is a real goal in nuclear engineering and materials science. To that end, the use of Molecular Dynamics (MD) simulations represent a unique tool, able to relate structural correlations and molecular behavior with thermodynamic properties of any material.

The phase diagram of the Li-Pb system shows limited solubility of Li in Pb and an eutectic at 0.17 at % Li, where congruent melting of the Li4%Pb solid solution and the B2 LiPb compound coexist.

Liquid binary alloys may show marked deviations from simple ideal mixtures. This fact is the rule, rather than the exception, when the electronegativity difference between the constituents, $\Delta\chi$, is large as in Li-Pb ($\chi_{Li}$ = 0.98, $\chi_{Pb}$ = 2.33 in Pauling scale). In this system the measured thermodynamic, electrical and magnetic properties exhibit pronounced deviations from ideality at specific stoichiometric compositions [3, 4]. Electron transfer from Li to Pb, gives rise to ionic bonding and preferred heterocoordination. Li-Pb systems belong to the class of alloys having a salt-like arrangement and no covalent bonding [5].

The work of Ruppersberg and Reiter [6] shows that Li–Pb alloys manifest a preference to an unlike atom arrangement leading to a negative short range order (SRO) in the liquid. A recent work shows that atomic distribution in the $Li_{17}Pb$ eutectic seems to be significantly affected by the $Li_4Pb$ associates [8]. These chemically ordered structural units (also called Zintl ions) are randomly distributed in the liquid Pb [8]. In the theoretical work of Ansionwu [9] the influence of the strong heterocoordination tendency of the LiPb liquid alloys is studied using a statistical thermodynamic model based on compound formation. In addition to the already mentioned $Li_4Pb$ compound, their study shows that the compound $Li_3Pb$ also has a profound influence on the thermodynamic properties of the liquid alloy.

Electrical and magnetic properties of liquid Li-Pb (and others alkali-group-IV liquid binary alloys) also exhibit pronounced deviations from ideality at specific stoichiometric compositions. A phenomenological theory [10] to explain the experimental data that has been somewhat successful assumes the existence of chemical complexes or aggregates with finite lifetimes, a concept lacking experimental support. However, recent *ab initio* MD simulations of liquid $Li_{50}Pb_{50}$ and $Li_{80}Pb_{20}$ alloys point to the fact that the chemical complex $Li_4Pb$ does not exist in the liquid Li–Pb alloys [11].

---

∗ Corresponding author. Tel.: +34-947-259-062; e-mail: albertofrailegarcia@gmail.com, scuesta@ubu.es

The study reported in this manuscript is, to our knowledge, *the first work in the literature attempting to model a complex system like LiPb at the atomic scale in liquid phase in a wide range of concentrations*, and capable of provide a good description of the complex phenomenology and properties of the Li-Pb alloys. To that end, we use Embedded Atom Method, EAM, potentials [13] since they are a balanced optimal choice between accuracy and computational cost to simulate metals in solid and liquid phases. Good predictions for the liquid phases of Pb and Li have been confirmed in our previous work [12] using the EAM potentials described in [14,15] for Pb and Li respectively. Based on them, we develop here an alloy potential aiming at describing the binary system. An evaluation of a number of liquid properties compared with available experimental data is presented certifying the validity of our development.

## 2. Computational details.

Molecular dynamics simulations have been carried out using the LAMMPS package [16]. Liquid samples are prepared from solid solutions run long enough to establish the equilibrium liquid structure (20 ps after melting). Size effects have been checked and our simulations with N = 15000 atoms are equivalent to those with N = 500000 atoms. Further details of the MD methods used can be found elsewhere [12].

Since the test of the quality of the published Pb and Li potentials proved to be satisfactory for the properties of the pure liquids, we used the effective representation (ER) to rescale both potentials [17]. The ER normalizes densities and pair potentials but does not alter the properties of the pure elements; it has the advantage of minimizing the contribution of the embedding term to the formation energy of the alloy, giving to the cross pair potential the task of creating most of the alloying effects. The ER allows us to combine potentials for pure elements coming from different procedures with eventually very different and unrelated magnitudes of the densities.

Several approaches could be followed to create a LiPb potential, such as fitting to *ab initio* calculations of key properties, force-matching method, lattice inversion etc. Here we decided to target the experimental mixing energy [18, 19]. This method implies an important simplification as is to assume that the (experimental) heat of formation of the liquid alloy is the same as in the solid phase at low temperature, which is used in the fitting procedure, an approximation whose accuracy is tested *a posteriori* by evaluating the predictions for the liquid phase properties.

We use the Redlich-Kister expansion to describe the mixing energies [20] with values for Li-Pb from [18, 19].

$$\Delta g(x,T) = x(1-x)\sum_{p=0}^{n} L_p(T)(1-2x)^p \qquad (2)$$

Where $L_p$ is the $p$th-order binary interaction parameter; in general, it is a function of temperature. We find that n=3 is adequate for this system.

To ensure that the potential gives correct thermodynamic properties we also use the cohesive energy, $E_c$, and lattice parameter of the B2 LiPb compound (the stable phase at high temperature) as targets in the fitting procedure. This compound has two different crystal structures (CS) in the solid phase [21]. At low temperature (T < 200ºC) CS is the "hexagonal" hR6 (space group, R-3m). At high temperature (200 < T < 482 ºC) CS is bcc ClCs-like, i. e. two simple cubic structures interpenetrating (Pearson Symbol cP2, space group Pm-3m).

The Li-Pb pair interaction is written in terms of a piece-wise cubic polynomial with nk knots at chosen positions ri, and the Heaviside step function H(x) as,

$$V_{LiPb}(r) = \sum_{i}^{n_k} a_i H(r_i - r)(r_i - r)^3 \qquad (3)$$

## 3. Results.

Note that different parameterizations can be obtained by giving different weights to $\Delta H(x)$ and $E_c$ . The influence on the final quality of a potential deserves a further discussion [22]. The potential presented here, M1, has been chosen as the compromise of parameters best reproducing physical and thermodynamic properties of the PbLi system.

Figure 1 shows the mixing enthalpy of Li-Pb at 1000 K. It is important to note that the fitting procedure is done for an unrelaxed 0 K alloy or compound (see Ref. [17]), and that the good agreement with experiments obtained when the simulation is done at 1000K reflects that the enthalpy of mixing is weakly dependent on T [18, 19]. Cohesive energy of B2 LiPb and lattice parameter are 2.415 eV/at [23] and 3.586 Å [21] at 0 K and 230 ºC respectively. Note that the compound is more stable than the solution by more than 0.2 eV/atom. This will turn important to have the correct melting temperature, as it will be further discussed below.

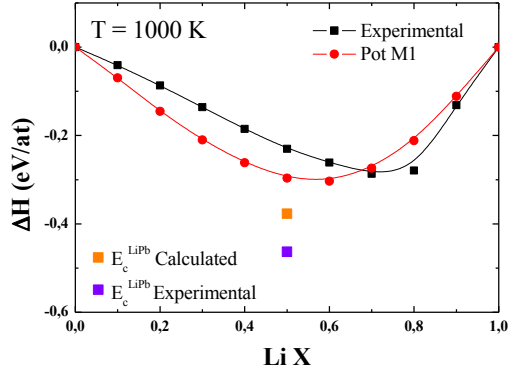

**Fig. 1**. (Colour online) Enthaphy of mixing at 1000 K vs Li concentration. Experimental values taken from [19], MD results as labelled. Cohesive energy of LiPb in B2 phase at 0 K, (purple square) is also used as a target. Orange square represent the result obtained with our LiPb potential.

The enthalpy of mixing in a disordered structure can be defined analytically, as well as the cohesive energy of the B2 phase knowing the CS, just expanding the energy expression in the EAM formalism to a certain sum of terms depending on the cutoff. To create the M1 potential we minimized the result obtained after adding two squared differences; first the enthalpy of mixing difference between the calculated one (ΔH(fcc)) and the experimental one (ΔH(exp)) calculated with our potential. Second the difference between the cohesive energy obtained with our potential at a certain lattice parameter (3.586 Å in our case) and the experimental one (-2.41 eV/at). To increase the importance of the 50-50 compound we add a factor 2 to weight it. This procedure is repeated iteratively till convergence is reached.
We defined a pair potential for Li-Pb interaction as a sum of spline knots.

We rewrite equation 2 for convenience as:

$$V_{LiPb}(r) = \sum_{i=1}^{n_k=4} c_i H(r_c - r_{x0i})(r_c - r_{x0i})^2 (r_c - r_{xbi})$$

Where $r_c$ is 6 Å. (is not the cutoff). In this way 12 parameters define our LiPb potential. (See table I). The potential file is available under request.

| $C_1$ | 0,01217 | $X_{b1}$ | 2.8354  | $X_{01}$ | -0.07122 |
|-------|---------|----------|---------|----------|----------|
| $C_2$ | 0,01215 | $X_{b2}$ | -9.0614 | $X_{02}$ | 2.5767   |
| $C_3$ | 0,00005 | $X_{b3}$ | -1166.9 | $X_{03}$ | 3.15786  |
| $C_4$ | 0,01    | $X_{b4}$ | 1.00    | $X_{04}$ | 6.5      |

Table I. The 12 parameters defining our LiPb potential as described in the text.

A detailed analysis of the properties predicted by our potential for different Li-Pb alloys confirms the validity of our contribution:

 - In Fig. 2 we present the structure factor q(S(q)-1) curves of liquid Li17Pb alloys at 775 K compared to the experimental data reported in [7]. MD results are shown and were calculated using the Debyer code [24] after MD runs with our Li-Pb potential.

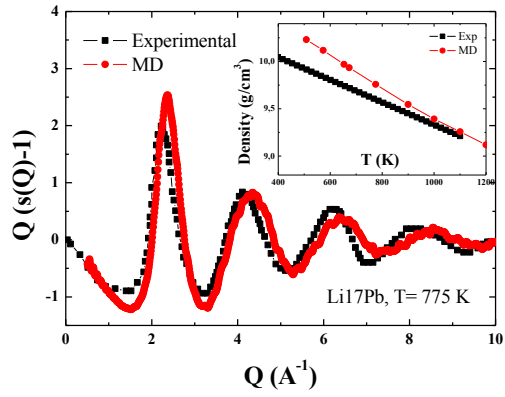

**FIG. 2.** (Color online). q(S(q)-1) curves of liquid Li17Pb alloys at 775 K. Black squares are the experimental data taken from [7] red circles represent the calculated curves from MD simulations. Inset shows density variation with temperature compared to experimental data as labeled.

Satisfactory agreement between the structure factor calculated by MD and the experimental one implies a similarly good agreement in atomic density. Indeed, our density calculations give very good results in a wide temperature range, from 500 K to 1200 K. (see Inset in Fig 2). It is to be noted that Li17Pb alloy has a Li concentration of less than 0.7 % in weight, so mass density is governed by the properties of the Pb potential. Similar results were found regarding heat capacity (Inset in Fig. 3), calculated as derivative of enthalpy at constant pressure.

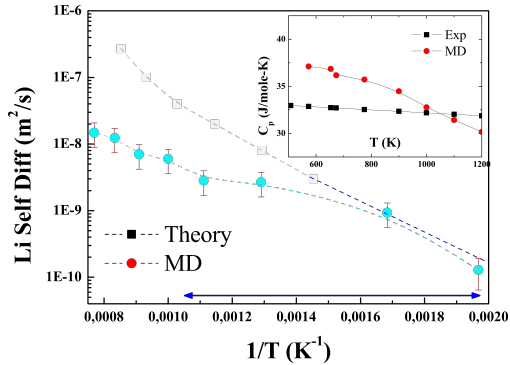

**FIG 3**. Lithium self-diffusion in liquid Li17Pb vs 1/T. Red symbols represent the values calculated from MD simulations (N =13500 atoms), in black squares the theoretical estimation taken from [26]. Blue arrow denotes the working temperature window in breeder blankets. The inset shows the calculated heat capacity compared to the experimental values according to [26].

Finally, another property of interest of the liquid is lithium self-diffusion at the eutectic concentration. The calculated values from MD simulations using the usual Einstein-Stokes formula [25] are compared in Fig. 3 to the theoretical estimation reported in [26]. Our values are of the same order of magnitude but present slightly different temperature dependence at higher temperatures.

Let's now examine the Li-Pb system at other concentrations. Structural properties of higher Li concentration alloys are controversial. Experimental X-Ray and neutron scattering results differ significantly; a detailed analysis can be found in reference [6]. Our MD results agree better the neutron scattering data.

Our calculation of S(q) does not make any difference between Li and Pb atoms (i.e. different scattering sections) and hence S(q) is not expected to match the experimental results shown in Fig. 2 of Ref [6].

For binary phases (i=1, 2), S(q) may be decomposed into the weighted sum of the Bhatia-Thornton [27] partial structure factors:

$$S(q) = w_{NN}S_{NN}(q) + w_{NC}S_{NC}(q) + w_{CC}S_{CC}(q) \quad (4)$$

where the weight factors are written in terms of the concentrations $c_1$, $c_2$ and the neutron scattering lengths $b_1$, $b_2$ as

$$w_{NN} = \frac{(c_1 b_1 + c_2 b_2)^2}{c_1 b_1^2 + c_2 b_2^2}, \quad w_{NC} = \frac{2(c_1 b_1 + c_2 b_2)(b_1 - b_2)}{c_1 b_1^2 + c_2 b_2^2}$$

$$\text{and} \quad w_{CC} = \frac{(b_1 - b_2)^2}{c_1 b_1^2 + c_2 b_2^2} \quad (5)$$

But coherent scattering cross sections (in barns) for pure Pb and Li (99.05% $^7$Li, 0.95% $^6$Li) are very different, $b_1 = \sigma_{Pb} = 11,11$ b and $b_2 = \sigma_{Li} = 0,66$ b [7].

For $b_1 = b_2$ the neutrons does not distinguish between particles 1 and 2, and one observes the global structure, which is related to $S_{NN}(q)$ since the other terms in equation (4) vanish. $S_{NN}(q)$ is the term governing density and is similar to that of a hard-sphere mixture [28]. Our numerical S(q) results, obtained without making any difference in the intensity of scattering of atoms, match rather well this $S_{NN}(q)$ term (see Fig 4). Hence a correct value of the density of liquid LiPb is obtained as shown in the inset of Figure 4.

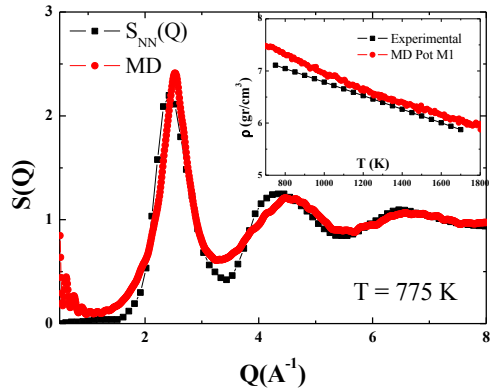

**FIG. 4**. LiPb alloys at 775 K; total structure calculated from MD simulations (red lines) compared to $S_{NN}(q)$ taken from [6]. Inset shows liquid LiPb density vs T; red circles represent the MD values and black squares are the experimental data [29]. Statistical errors are the size of the symbols and are not shown for clarity.

Is also noteworthy that $S_{NN}(q)$ is very similar to S(q) calculated in a HS approximation [6] what can be explained since pure Pb behaves as a liquid of HS according to [28]. Thus, LiPb system ($m_{Pb} \sim 30 m_{Li}$) seems governed by the heaviest atom (Pb) when Li concentration is low (eutectic for example) and up to 50% of Li concentration, presenting a HS structure. This is not surprising since even in the 50-50 concentration Li represents less than 4 % of the total mass. Our calculated S(q) is close to $S_{NN}(q)$ implying that calculated densities are close to the experimental values.

To support this affirmation, in Figure 5 we present the pair distribution function of liquid Li-Pb alloys at 1000 K calculated from MD simulations. As can be seen g(r) does not change much going from pure Pb to high Li concentrations, just a smooth shift of the first and second peak towards lower radii is observed.

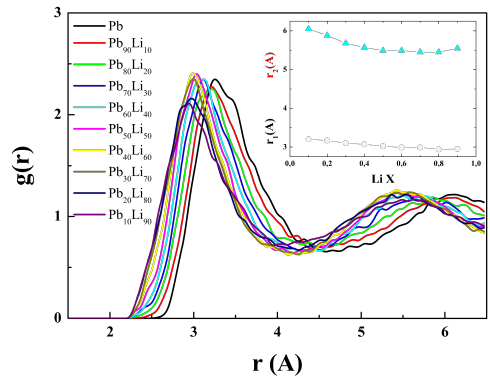

**FIG. 5.** Pair distribution function, g(r), of liquid Li-Pb alloys at 1000 K. Inset shows the variation of $r_1$ (black circles) and $r_2$ (red triangles) with Li concentration.

Another physical magnitude to analyze providing relevant structural information is the short range order (SRO). Experiments using neutron diffraction show SRO values of -0.25 for liquid Li-Pb alloys ($Li_{80}Pb_{20}$, $Li_{62}Pb_{38}$ and $Li_{50}Pb_{50}$) [6] while in our MD simulations the mixtures are almost random distributions, with slightly negative Warren-Cowley [29] SRO values (see Fig. 10). It has to be noted that the authors admit: "[SRO] *numerical values, especially for the non-stoichiometric alloys, have been obtained in a quite speculative way*" [6].

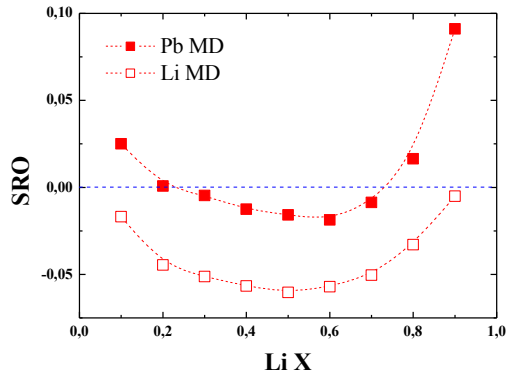

**FIG. 6**. (Colour online) SRO values for Li-Pb system using EAM/alloy potential. Closed symbol correspond to Pb SRO values and open symbols to Li SRO ones.

Negative SRO is in agreement with the existence of an ordered solid phase, and also in agreement with the fact that this system has a negative heat of formation,
(See Fig. 1).

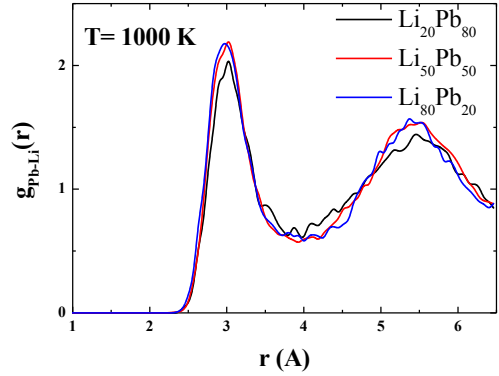

**FIG. 7**. (Colour online) Partial $g_{Li-Pb}(r)$ curves obtained from MD simulation for three different Li-Pb alloys as labeled. As can be seen no clear heterocoordination tendency is observed moving from low to high Li concentrations.

Figure 7 shows the partial $g_{Li-Pb}(r)$ for $Li_{20}Pb_{80}$, $Li_{50}Pb_{50}$ and $Li_{20}Pb_{20}$ at 1000 K. The number of Li-Pb pairs does not increase so much going from low to high Li concentrations. It has to be noted that our MD simulations give very similar results to that obtained *ab initio* by Senda *et al* [10]. First peak positions are almost the same in our classical MD and second peak positions only differ slightly to that presented in Ref. [10].

The theoretical work of Hafner and coworkers [30] shows how the concept of chemical complexes can be abandoned when using appropriate interatomic potentials. In short they model liquid Li-Pb alloys with a set of potentials containing a HS potential for the mean pair interaction, $\varphi_{NN}$, (see [30] for details) and a HS potential of the same diameter plus a Yukawa tail for the ordering potential, $\varphi_{CC}$. They found good agreement to different structural properties. We also find a volume contraction at the same concentration that observed experimentally.

Is not our aim to compare our model to the theoretical one presented in [30] but is to be noted that, similar to our simulations, these studies prove that different deviations in some properties (volume and heat capacity) can be observed without introducing of $Li_4Pb$ or $Li_3Pb$ complexes.

Moreover, our simulated liquid Li-Pb system still presents some complex features. Experimentally, a volume contraction is observed around 80 % of Li [31] sometimes claimed to be due to some kind of $Li_4Pb$ compound formation. In figure 6 we plot liquid $Pb_xLi_{1-x}$ volume, V(x), (at 1000 K) vs concentration, x, and a clear volume contraction is shown around the expected value of Li concentration.

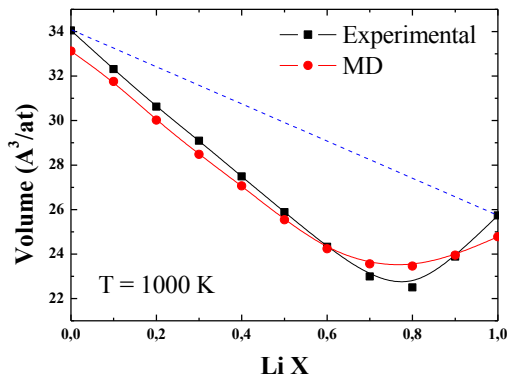

**FIG. 8.** Liquid Li-Pb alloys at T = 1000 K. Experimentally, volume shows a departure from ideality (blue dashed line) probably due to compound formation. MD results also present similar volume contraction around $Li_{80}Pb_{20}$.

Not only the volume presents a departure of linear behavior, the same can be observed in the variation of density with temperature, $d\rho/dT$, in total agreement with experiments [31]. Results are presented in Figure 9.

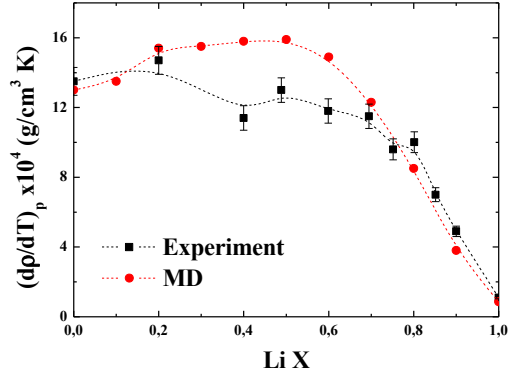

**FIG. 9**. Density variation with temperature of liquid Li-Pb alloys compared to experimental results [31]. MD results calculated at 1000 K. Statistical error are about the size of the markers and not shown for the shake of clarity.

The results presented in Figure 9 will be important in what follows. It is known that heat capacity can be written as [29]

$$C_p = -MT\left(\partial p/\partial T\right)_s \left(\partial \rho/\partial T\right)_p \qquad (6)$$

where M stands for the molar weight. Hence, the density variation with temperature is well described by our potential (as can be seen in Fig. 9) while we do not know how $(\partial p/\partial T)_s$ will be.

Also related to thermodynamic properties, another important deviation from ideal solution is observed in the heat capacity as Li concentration is varied, $C_p$, [31]. Figure 10 shows our $C_p$ results compared to the experimental data reported by [31]. Interestingly, our interatomic potential shows concentration dependence very similar to the experimental curve. The electronic contribution is small, almost negligible for Li x < 50 and about 5 J/mole-K for high Li concentrations [32].

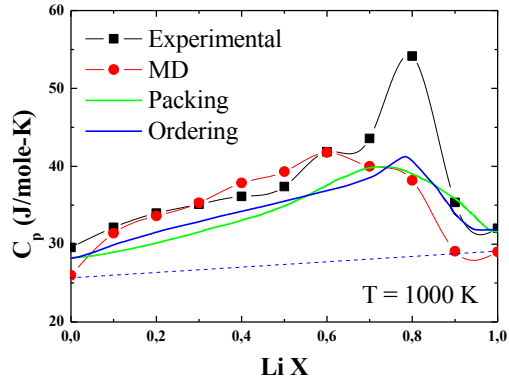

**FIG. 10**. Liquid Li-Pb alloys at T = 1000 K. Experimentally [31], heat capacity present a strong departure from ideality (violet dashed line). MD results also present similar behaviour. Blue and green lines represent the calculated $C_p^{ord}$ and $C_p^{pack}$ contributions calculated in [32].

In our MD simulations the $(d\rho/dT)_p$ contribution is only slightly underestimated for high Li concentrations and hence it is not responsible of the observed difference in $C_p(x)$. Contrary, the $(dp/dT)_s$ has to be underestimated in our MD simulations in order to give us the results presented in Fig 10.

We can rewrite $(dp/dT)_s$ like

$$\left(\frac{\partial p}{\partial T}\right)_s = \left(\frac{\partial p}{\partial \rho}\right)_s \left(\frac{\partial \rho}{\partial T}\right)_s = c^2 \left(\frac{\partial \rho}{\partial T}\right)_s \qquad (7)$$

where c is the velocity of sound by definition. Hence it is noteworthy that the reason of the excess heat capacity that is not correctly described by our simulations seems to be related with c, a quantity that it is known to present interesting peculiarities around the 80% of Li concentration due to collective excitations sometimes referred as "fast sound" [33]. Indeed, the existence of such excitations was suggested on the basis of MD simulations of molten $Li_4Pb$ [33]. Some years later INS experiments conducted on molten $Li_4Pb$ and $Li_4Tl$ [34] obtained the first experimental evidence of the existence of collective modes with a phase speed much higher than the hydrodynamic speed of sound in liquid metallic alloys*.

It has to be noted that our MD results are quite similar to the calculated order and packing contribution to the excess heat capacity as presented in [32] (see Fig 10). Ruppersberg and Saar [32] calculated the heat capacity of Li-Pb (and Na-Pb and K-Pb) using the Hafner-Pasturel-Hicter formalism [30]. They used the ordering potential at contact as an adjustable parameter fitted to the values of $\Delta H_{ord}$ defined as $\Delta H_{ord}=\Delta H-\Delta H_{eg}$. where '*eg*' stands for electron gas. According to [30] the electron gas contribution calculated within the framework of the Pines-Nozieres approximation [36] to electronic exchange contribution to the total ΔH is quite high (about a 40 % for high Li concentrations), but that calculation could be quite overestimated since the electron contribution to the total energy is about a 25 % for pure Pb and much less than a 5 % in pure Li [37].

Last, we report on solid and liquid free energy calculations for LiPb in B2 phase following the switching Hamiltonian method as described in [38, 39]. The results represent another proof of the suitability of our potential; calculated melting of LiPb, $T_m(MD) = 720 \pm 20$ K, is very close to the experimental value at 755 K [31].

*Note: The dynamics in liquid binary mixtures with different mass ratio of components has been recently studied with the generalized collective modes approach. T. Bryk and I. Mryglod. Collective dynamics in binary liquids: spectra dependence on mass ratio. *J. Phys.: Condens. Matter* 17 413 (2005) [35].

## 4. Conclusions

In summary, we have developed a new Li-Pb interatomic potential that correctly describes the properties of eutectic Li17Pb. Our MD simulations gave good match of many structural, thermodynamic and dynamic properties around the eutectic concentration, of crucial importance in current nuclear fusion developments. This interatomic potential is the first step towards the creation of a ternary potential Li-Pb-He that will allow us to simulate He bubble cavitation and nucleation processes which are crucial phenomena in blanket designs for Fusion prototypes [1, 2].

Higher Li concentrations, as expected, showed more difficulty to be described by a simple EAM potential due to possible charge transfer between Li and Pb ions. Structural properties of the simulated LiPb are similar to those of a hard sphere mixture. Nevertheless we have shown that an EAM potential can give good results, in a complex compound forming system like Li-Pb, regarding its volume contraction and excess heat capacity. Interestingly, the excess heat capacity observed in Li-Pb alloys around the 80% of Li concentrations seems related to collective dynamics. However, to analyze in deep the dynamics of Lithium rich Li-Pb alloys is beyond the scope of our work.

On the technological side, this work sets the basis for a multiscale modeling tool aiming to shed light on the complex phenomena of helium transport in liquid metals (i.e. Li17Pb, Li). Moreover, it is going to provide data about the behaviour of the LiPb eutectic (Li 17%) in temperature regions where experimental results are still not available. Both advances are of critical importance in future nuclear technology designs.

Also important, our simulations re-open the debate about the structure and dynamics of liquid LiPb alloys. New neutron and X-ray diffraction experiments are necessary to fully understand the structure of liquid Li-Pb system.

## 5. Acknowledgments

This work is partially funded by the European CONSOLIDER Program. The work of the first author is part of their PhD Thesis and has been supported by the Universidad Politécnica de Madrid (Spain). We are grateful to Professor D. Belaschenko for providing us with his Li potential and many useful discussions. We acknowledge L. Zhang for interesting discussions and assistance with SRO calculations and N. Gupta for assistance in some Mathematica calculations.

**References and notes**

[1] Wong C. P. C. An overview of dual coolant Pb–17Li breeder first wall and blanket concept development for the US ITER-TBM design. Fusion Engineering and Design 81 (2006) 461–467.
[2] Norajitra P. The EU advanced dual coolant blanket concept. Fusion Eng. Des. 61–62 (2002) 449–453.
[3] Chieux P. and Ruppersberg H. 1980 *J. Physique Coli.* 41 C8 145-52
[4] van der Lugt W. and Geertsma W. 1984 *J. Non-Cnst. Solids* 61-2 187-200
[5] W. Geertsma et al. Electronic structure and charge-transfer-induced cluster formation in alkali-group-IV alloys. J. Phys. F: Met. Phys. 14 (1984) 1833-1845.
[6] Ruppersberg H. and Reiter H. J. Phys. F:Met. Phys. 12 1311 (1982).
[7] Ruppersberg H. and Egger H. 1975 *J. Chem. Phys.* 63 4095.
[8] Mudry S. *et al.* Journal of Nuclear Materials 376 (2008) 371–374.
[9] Anusionwu B.C. *et al.* Physics and Chemistry of Liquids Vol. 43, No. 6, (2005), 495–506
[10] Y. Senda, F. Shimojo and K. Hoshino. The ionic structure and the electronic states of liquid Li–Pb alloys obtained from *ab initio* molecular dynamics simulations. J. Phys.: Condens. Matter 12 (2000) 6101–6112.
[11] Hoshino K and Young W H 1980 *J. Phys. F: Met. Phys.* 10 1365-74
[12] Fraile A, Cuesta-López S, Caro. A, Perlado J. M. Journal of Nuclear Materials Vol 440, 1–3, September 2013, Pages 98–103
[13] Baskes M. I. (1988), MRS Bulletin, 13, 28. Foiles S. M. Embedded-atom-method functions for the fcc metals Cu, Ag, Au, Ni, Pd, Pt, and their alloys. Phys. Rev. B 33, 7983 (1986).
[14] Zhou X. W. Atomic scale structure of sputtered metal multilayers. Acta Mater. 49, 4005 (2001).
[15] D. Belashchenko et al. High Temperature 2009 vol 47 No 2 211-218. D. K. Belashchenko. Inorganic Materials, Vol.48, No. 1, pp. 79–86 (2012).
[16] Plimpton S. Fast Parallel Algorithms for Short-Range Molecular Dynamics, J Comp Phys, 117, 1-19 (1995).
[17] Caro A. et al. Phys Rev Lett95, 075702 (2005)
[18] W. Gasior and Z. Moser, Journal of Nuclear Materials 294 (2001) 77-83
[19] H. Ruppersberg and W. Speicher. Z. Naturf: 31 47-52 (1976) (taken from Hafner 1984, Ref [33])
[20] Saunders N. and Miodownik A. P., CALPHAD: A Comprehensive Guide, edited by R.W. Cahn, Pergamon Materials Series (Oxford, New York, 1998).
[21] Zalkin A. and Ramsey W.J . *Phys. Chem.*,1957, 61 (10), pp 1413–1415. Zalkin A. and Ramsey W. J. 1958 J. Chem. Phys. 62 689-93.
[22] Fraile A, Cuesta-López S, Caro. A. *et al.* To be published.
[23] L. M. Molina, J. A. Alonso and M. J. Stott. Solid State Communications. Vol. 108, Issue 8, 20(1998), pp 519-524
[24] http://code.google.com/p/debyer/
[25] J.P. Hansen, I.R. McDonald, Theory of Simple Liquids, Academic Press, New York, 1986.
[26] Mas de les Vall E. Lead-lithium eutectic material database for nuclear fusion technology. Journal of Nuclear Materials 376 (2008) 353-357R.
[27] A. B. Bhatia and D. E. Thornton. Phys. Rev. B 1 3004-12 (1970)
[28] Waseda Y. The Structure of Non-Crystalline Materials: Liquids and Amorphous Solids (McGraw-Hill, NewYork, 1980).
[29] Warren B.E.,X-Ray Diffraction (Addison-Wesley, Reading M A, 1969).Cowley J.M., Phys. Rev. 77, 667 (1950).
[30] J. Hafner, A. Pasture and P. Hicter. Simple model for the structure and thermodynamics of liquid alloys with strong chemical interactions. I: Chemical short-range order. J. Phys. F: Met. Phys. 14 (1984) 1137-1 156.
[31] J. Saar and H. Ruppersberg. 1987 J. Phys. F: Met. Phys. 17 305
[32] H. Ruppersberg and J. Saar. Calculation of the heat capacity of liquid Pb/Li, Na, K alloys according to the HPH formalism. 1989 *J. Phys.: Condens. Matter* 1 9729.
[33] J. Bosse, G. Jacucci, M. Ronchetti, and W. Schirmacher, Phys. Rev. Lett. 57, 3277 (1986).
[34] P. H. K. de Jong, P. Verkerk, C. F. de Vroege, L. A. de Graaf, W. S. Howells, and S. M. Bennington, J. Phys.: Condens. Matter 6, L681 (1994).
[35] T. Bryk and I. Mryglod. Collective dynamics in binary liquids: spectra dependence on mass ratio. *J. Phys.: Condens. Matter* 17 413 (2005).
[36] Pines D and Nozieres Ph 1966 *The Theory of Quantum Liquids* (New York: Benjamin)
[37] A. M. Vora. Thermodynamic properties of liquid metals using Percus–Yevick (PY) hard sphere reference system. *Journal of Engineering Physics and Thermophysics* 82:4, 779-788. Online publication date: 1-Jul-2009.
[38] E. M. Lopasso, M. Caro, A. Caro, and P. E. A. Turchi. Phys. Rev. B 68, 214205 (2003)
[39] Ciccotti G. and Hoover W. G. Molecular-Dynamics Simulation of Statistical-Mechanical Systems, (North-Holland, Amsterdam, 1986), pp. 169–178.